\documentclass[aps,prb,twocolumn,superscriptaddress]{revtex4}
\usepackage{graphicx}

\begin{document}

\title{Short-range cluster spin glass near optimal superconductivity
in BaFe$_{2-x}$Ni$_{x}$As$_{2}$}
\author{Xingye Lu}
\affiliation{Beijing National Laboratory for Condensed Matter
Physics, Institute of Physics, Chinese Academy of Sciences, Beijing
100190, China}
\author{David W. Tam}
\affiliation{ Department of Physics and Astronomy,
Rice University, Houston, Texas 77005, USA }
\author{Chenglin Zhang}
\affiliation{ Department of Physics and Astronomy,
Rice University, Houston, Texas 77005, USA }
\author{Huiqian Luo}
\affiliation{Beijing National Laboratory for Condensed Matter
Physics, Institute of Physics, Chinese Academy of Sciences, Beijing
100190, China}
\author{Meng Wang}
\affiliation{Beijing National Laboratory for Condensed Matter
Physics, Institute of Physics, Chinese Academy of Sciences, Beijing
100190, China}
\author{Rui Zhang}
\affiliation{Beijing National Laboratory for Condensed Matter
Physics, Institute of Physics, Chinese Academy of Sciences, Beijing
100190, China}
\author{Leland W. Harriger}
\affiliation{NIST Center for Neutron Research, National Institute of
Standards and Technology, Gaithersburg, Maryland 20899, USA}
\author{T. Keller}
\affiliation{Max-Planck-Institut f$\ddot{u}$r Festk$\ddot{o}$rperforschung, Heisenbergstrasse 1, D-70569 Stuttgart, Germany}
\affiliation{Max Planck Society Outstation at the Forschungsneutronenquelle Heinz Maier-Leibnitz (MLZ), D-85747 Garching, Germany}
\author{B. Keimer}
\affiliation{Max-Planck-Institut f$\ddot{u}$r Festk$\ddot{o}$rperforschung, Heisenbergstrasse 1, D-70569 Stuttgart, Germany}
\author{L.-P. Regnault}
\affiliation{SPSMS-MDN, UMR-E CEA/UJF-Grenoble 1, INAC, Grenoble, F-38054, France}
\author{Thomas A. Maier}
\affiliation{Center for Nanophase Materials
Sciences and Computer Science and Mathematics Division, Oak Ridge National Laboratory, Oak Ridge, Tennessee 37831-6494, USA}
\author{Pengcheng Dai}
\email{pdai@rice.edu}
\affiliation{ Department of Physics and Astronomy,
Rice University, Houston, Texas 77005, USA }
\affiliation{Beijing National Laboratory for
Condensed Matter Physics, Institute of Physics, Chinese Academy of
Sciences, Beijing 100190, China}



\begin{abstract}
High-temperature superconductivity in iron pnictides occurs when
electrons are doped into their antiferromagnetic (AF) parent compounds.
In addition to inducing superconductivity, electron-doping also changes the static commensurate AF order in the
undoped parent compounds into short-range incommensurate AF order near optimal superconductivity. Here we use neutron scattering to demonstrate
that the incommensurate AF order in BaFe$_{2-x}$Ni$_{x}$As$_{2}$ is not a spin-density-wave arising
 from the itinerant electrons in nested Fermi surfaces, but consistent with a cluster spin glass
 in the matrix of the superconducting phase.  Therefore, optimal superconductivity in iron pnictides
 coexists and competes with a mesoscopically separated cluster spin glass phase, much different from the homogeneous coexisting AF and superconducting phases in
 the underdoped regime.
\end{abstract}

\maketitle

\section{Introduction}
A complete determination of the structural and magnetic phases in solids forms the basis for a comprehensive understanding of
their electronic properties \cite{kamihara,cruz,sefat,qhuang}.  For iron pnictides such as BaFe$_2$As$_2$, where superconductivity can be induced
by electron-doping via Co or Ni substitution, 
extensive transport \cite{canfield,jhchu} and neutron diffraction work \cite{hayden,pratt1,Christianson,mywang} have established the overall structural
and magnetic phase diagrams for BaFe$_{2-x}$Co$_{x}$As$_{2}$ \cite{nandi,pratt2} and BaFe$_{2-x}$Ni$_{x}$As$_{2}$ \cite{hqluo,xylu}.
In the undoped state, BaFe$_2$As$_2$ forms a collinear antiferromagnetic (AF) order with moment along the $a_o$-axis direction of the orthorhombic structure
[see left inset in Fig. 1(a)] \cite{qhuang}.  Upon electron doping to induce superconductivity, the static ordered moment and
 the N${\rm \acute{e}}$el temperature
($T_N$) of the system decreases gradually with increasing $x$ \cite{hayden}.  While the static AF order is commensurate with the underlying lattice and
coexists with superconductivity in the underdoped regime \cite{pratt1,Christianson,mywang}, it abruptly changes into transversely incommensurate short-range order
for $x$ near optimal superconductivity \cite{pratt2,hqluo,xylu}.  This has been hailed as direct evidence that the static AF order in iron pnictides
arises from the formation of a spin-density wave driven by itinerant electrons and Fermi surface nesting of the electron and hole pockets \cite{pratt2,jdong}, much like
the spin-density wave state of the chromium alloys \cite{fawcett}.

If the incommensurate AF order in iron pnictides near optimal superconductivity indeed arises from
the itinerant electrons and nested Fermi surfaces, one would expect that its incommensurability $\delta$ near
the AF ordering wave vector, or ${\bf Q}=(1,\pm\delta,3)$ in
the right inset of Fig. 1(a), will increase smoothly with increasing electron-doping due to the gradually mismatched
electron and hole Fermi surfaces [Figs. 1(b)-1(e)]. Systematic neutron diffraction experiments on BaFe$_{2-x}$Co$_{x}$As$_{2}$ \cite{nandi} and BaFe$_{2-x}$Ni$_{x}$As$_{2}$ \cite{hqluo,xylu} instead reveal a first-order-like commensurate to incommensurate transition with increasing electron-doping and a dramatic reduction in ordered moment in the
incommensurate phase.  Furthermore,  recent nuclear magnetic resonance (NMR) data on BaFe$_{2-x}$Co$_{x}$As$_{2}$ near optimal superconductivity indicate the presence of inhomogeneous
frozen AF domains (termed cluster spin glass) in the matrix of the superconducting phase \cite{dioguardi}.  If the short-range incommensurate AF order
in electron-doped iron pnictides \cite{pratt2,hqluo} is indeed the cluster spin glass phase, it cannot originate from the itinerant electrons in nested Fermi surfaces but may be a consequence of the
disordered localized moments \cite{ea,mydosh}.  Given the ongoing debate concerning the itinerant \cite{jdong} or localized \cite{qsi,cfang,ckxu} nature
of the antiferromagnetism in iron pnictides \cite{dai}, it is important to determine the microscopic origin of the incommensurate
AF order and its connection with superconductivity \cite{mazin,chubkov}.

\begin{figure}[t]
\includegraphics[scale=.45]{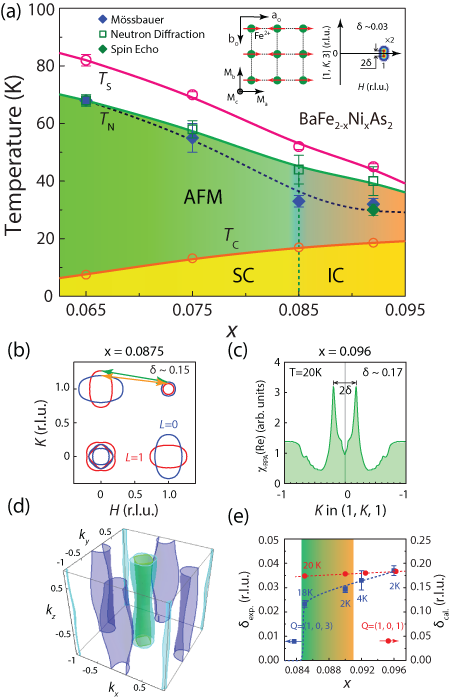}
\caption{(a) Electronic phase diagram of BaFe$_{2-x}$Ni$_x$As$_2$ in the underdoped regime.
The structural ($T_s$) and magnetic phase transitions ($T_N$) are taken from  Ref. \cite{xylu}.
Filled blue diamonds indicate the measured $T_N$ from M$\rm \ddot{o}$ssbauer measurements of the same samples \cite{munevar}.
The filled green diamond marks the $T_N$ from neutron resonance spin echo (NRSE) measurements. The left panel of the inset shows
the spin arrangement of iron in the AF ordered state. $M_a$, $M_b$ and $M_c$ are
the components of the ordered magnetic moment along the $a_o$, $b_o$, and $c$, respectively.
The right panel shows the incommensurate AF peaks in reciprocal space. The scale of the incommensurability is
multiplied by a factor of 2 for clarity. (b) Schematic diagram for Fermi surfaces and possible nesting wave vectors
for the $x=0.0875$ sample. The Fermi surfaces at $L=0$ and 1 are marked as blue and red, respectively.
The arrows indicate nesting wave vectors connecting $L=0$ and 1 planes.
(c) Wave vector dependence of the calculated RPA susceptibility at 20 K for the $x=0.096$ sample showing transverse incommensurability.
(d) Three dimensional Fermi surfaces of the system in reciprocal space.
(e) The electron doping evolution of the incommensurability from neutron diffraction experiments (blue squares) \cite{hqluo,xylu}
and RPA calculation (red circles). While the actual electron-doping levels are used in the RPA calculation, $x$ in the figure 
represents the nominal doping level for easy comparison with experiments.  
The colored region indicates mixed
phases where commensurate and incommensurate AF orders coexist. 
 }
\end{figure}

For a prototypical spin glass such as Cu$_{1-y}$Mn$_y$ alloy, the ordering temperature of the elastic magnetic scattering decreases systematically with increasing instrumental
energy resolution used to separate the true elastic component from the inelastic/quasielastic scattering \cite{murani}.
The magnetic order parameter is then the ``Edwards-Anderson'' order parameter \cite{ea}
measured with the spin relaxation time
$\tau\sim\hbar/\Delta E$ ($\Delta E$ is the neutron spectrometer energy resolution) below which the spins freeze \cite{murani}.
By using neutron spectrometers with vastly different energy resolutions ($1\ \mu {\rm eV}\leq\Delta E\leq1.5$ meV), we find that the ``N$\rm \acute{e}$el'' temperature
 of the incommensurate AF order in electron-doped BaFe$_{1.908}$Ni$_{0.092}$As$_{2}$ \cite{hqluo,xylu}
decreases from $T_N=36\pm 3$ K measured with $\Delta E\sim 1$ meV to $T_N=30\pm 2$ K for $\Delta E=1\ \mu$eV.  Furthermore, our polarized
neutron diffraction measurements indicate that the ordered moment direction of the incommensurate AF phase
 is along the longitudinal direction with no measurable component along the transverse direction.
By considering several possibilities for the incommensurate AF order, we conclude that it is a cluster spin glass (or more precisely moment amplitude spin glass)
in the matrix of the superconducting phase, coexisting and
competing with superconductivity \cite{dioguardi}.

\section{Theoretical calculation and Experimental results}

We carried out neutron scattering
experiments on BaFe$_{2-x}$Ni$_x$As$_2$
using SPINS, IN22, and TRISP triple axis spectrometers at
the NIST Center for Neutron Research, Institut Laue-Langevin, and MLZ,
respectively. Our samples are grown by the self-flux method \cite{ycchen}.  
From the inductively coupled plasma (ICP) analysis of the as-grown single crystals, we find that 
the actual Ni level is 80\% of the nominal level $x$ \cite{ycchen}. To allow direct comparison with our earlier measurements, we denote Ni-doping levels as the nominal level. 
For SPINS measurements, we used two-axis mode with incident beam energy of $E_i=5$ meV and
triple-axis mode with out-going neutron energy of $E_f=2.5$ meV.  
For the two-axis measurements, there is no analyzer in the exit neutron beam and neutron energies less than 5 meV can in principle be detected, yielding $0\leq \Delta E\leq 5$ meV.
For the triple-axis measurements, we have $\Delta E\approx 0.1$ meV.  For the TRISP measurements,
we used  $E_f=14.68$ meV and a 60 mm-thick pyrolytic graphite filter to remove $\lambda/2$
neutrons.  The instrument energy resolutions are $\Delta E\approx 1$ meV in the triple-axis mode
and $\Delta E\approx 1\ \mu$eV in the neutron resonance spin echo (NRSE) mode \cite{trisp,inosov}.  Finally, IN22 triple axis spectrometer in the polarized neutron
scattering mode was used to determine the moment direction of the incommensurate AF order with instrument setup described in Ref. \cite{hqluo13}.
  We define the wave vector $Q$ at ($q_x$, $q_y$, $q_z$) as $(H,K,L) = (q_xa_o/2\pi,
q_yb_o/2\pi, q_zc/2\pi)$ reciprocal lattice units (rlu) using the orthorhombic unit cell suitable for the AF ordered
iron pnictide, where $a_o\approx b_o \approx 5.6$ \AA\ and $c = 12.9$ \AA. In the undoped state, the commensurate AF order occurs at
${\bf Q_{AF}}=(1,0,L)$ with the ordered moment along the
$a_o$ direction of the orthorhombic unit cell [left inset in Fig. 1(a)] and $L=1,3,\cdots$ \cite{qhuang}.
For the experiments, we have used single
crystals of BaFe$_{2-x}$Ni$_x$As$_2$ with $x=0.092, 0.096$ where the incommensurate AF order
was found along the transverse direction at ${\bf Q}=(1,\pm \delta,3)$ [Fig. 1(a)] \cite{hqluo}.

\begin{figure}[t]
\includegraphics[scale=.45]{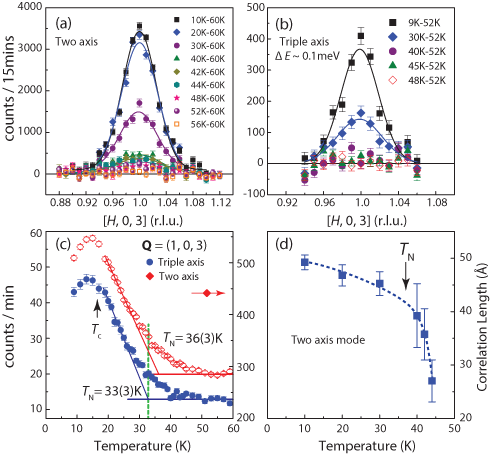}
\caption{
Temperature and wave vector dependence of the incommensurate AF ordering obtained on SPINS using two-axis and triple-axis modes
for the $x=0.092$ sample. (a) Temperature differences of the longitudinal scans
along the $[H,0,3]$ direction using the $T=60$ K data as background scattering. (b) Identical scans using
triple-axis mode with $\Delta E=0.1$ meV.
The solid lines are Gaussian fits to the data. (c) Comparison of the AF
order parameters between the two-axis and triple-axis measurements. The $T_N$ and $T_c$ are
marked by the arrow and intersects of the solid lines, respectively.
(d) The temperature evolution of the spin-spin
correlation length from the two-axis measurements with $T_N$ marked by the arrow.
The blue dashed line is a guide to the eye.
 }
 \end{figure}

Figure 1(a) shows the
electronic phase diagram of BaFe$_{2-x}$Ni$_x$As$_2$ as a function
of the nominal Ni-doping level $x$, where the commensurate to incommensurate AF phase
transition occurs around $x=0.085$ in the first order fashion \cite{hqluo,xylu}.
We first consider if the observed incommensurate AF order can be understood within the itinerant electron
Fermi surface nesting picture \cite{pratt2}.
Using a random-phase approximation (RPA) approach \cite{maier,hqluoprb12},
we calculate the magnetic susceptibility from a tight-binding 5-orbital Hubbard-Hund Hamiltonian
fitted to the density function theory (DFT) band structure for BaFe$_2$As$_2$ with a rigid band shift applied to account for electron doping \cite{graser}.
For the calculation, we assumed that only 80\% of the additional Ni electrons are doped into the Fe-As planes from the ICP measurements \cite{ycchen}.
This allows us to compare directly the calculated Fermi surface nesting wave vectors with the neutron scattering experiments.  
The interaction matrix in orbital space contains on site matrix elements for the intra- and inter-orbital
Coulomb repulsions $U$ and $U^\prime$, and for the Hunds-rule coupling
and pair-hopping terms $J$ and $J^\prime$. 
From the earlier work \cite{hqluoprb12}, we know that the RPA calculated Ni-doping evolution of the low-energy  
spin excitations 
is in qualitative agreement with the neutron scattering experiments.
To calculate the electron doping evolution of the incommensurate AF order, we
have used the spin rotationally invariant interaction parameters $U=0.8$, $U^\prime=U/2$, $J=U/4$, and $J^\prime=U/4$ well below the RPA instability threshold.
Figure 1(b) shows the Fermi surfaces of BaFe$_{2-x}$Ni$_x$As$_2$ at $x=0.0875$ and the arrows indicate the nesting wave vectors
between the hole pockets at $M$ point and electron pockets which gives the incommensurability
$\delta\approx 0.17$ [Fig. 1(c)].  Figure 1(d) shows the full three-dimensional Fermi surfaces used in the calculation.
While the calculated evolution of the Fermi surfaces with increasing electron-doping is qualitatively consistent
with those determined from angle resolved photoemission experiments \cite{richard}, 
comparing the electron doping dependence of $\delta$ from the RPA and
experiments [Fig. 1(e)] reveals that the RPA values of $\delta$ are about 5 times larger than the measured values
and do not exhibit the commensurate to incommensurate AF order transition near $x=0.085$ \cite{hqluo}.
Although our calculation does not include the electron-lattice coupling effect on 
the incommensurate AF order, we do not expect that including such an effect will induce first order like commensurate to incommensurate AF 
transition as a function of increasing Ni-doping. 
Therefore, the incommensurate AF order may not originate from a
spin-density wave in nested Fermi surfaces.

\begin{figure}[t]
\includegraphics[scale=.35]{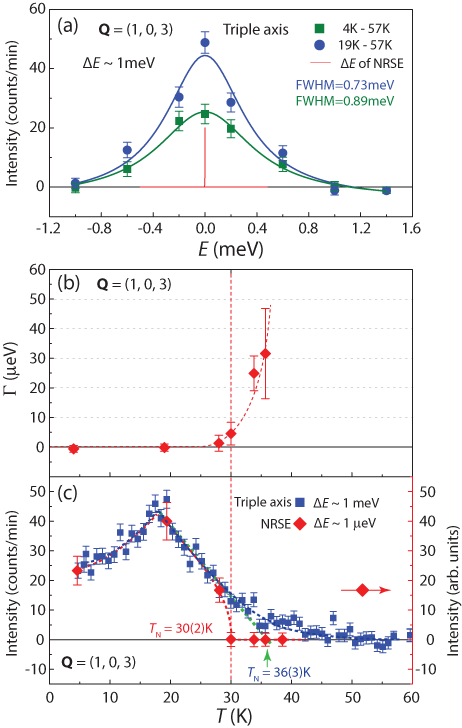}
\caption{Measurements on the $x=0.092$ sample using the TRISP triple-axis spectrometer.
(a) Temperature differences energy scans measured with the normal triple-axis spectrometer mode
at ${\bf Q} = (1,0,3)$.  They are instrumental resolution limited at $T=4$, and 19 K using the
$T=57$ K data as background. The red line indicates the effective energy resolution of the NRSE measurement at $\Delta E\approx 1\ \mu$eV.
(b) Evolution of the energy width with increasing temperature
at ${\bf Q} = (1,0,3)$. $\Gamma$ is the Half-Width-at-Half-Maximum (HWHM) of scattering function and zero indicates instrumental resolution limited.
(c) The Magnetic order parameters from the normal triple-axis (blue squares) and NRSE (red diamonds)
 measurements.
The reduction in intensity below $\sim$19 K is due to superconductivity \cite{hqluo}.
The $T_N$ from normal triple-axis is estimated by the intersect of two straight lines (green dashed line and black solid line)
from the linear extrapolation of the low and high-temperature data.
From NRSE data, one can determine the energy width and integrated intensity of the scattering function $S(Q,E)$.
The magnetic order parameter with $\Delta E=1\ \mu$eV
is shown in Fig. 3(c) with $T_N$ marked as vertical dashed line. The red dashed curves are guides to the eye.
 }
\end{figure}

To test if the incommensurate AF order arises from a cluster spin-glass as suggested from the NMR experiments \cite{dioguardi}, we
carried out neutron diffraction measurements using SPINS with two-axis and triple-axis modes \cite{clzhang}.  Figure 2(a) shows longitudinal
scans along the $[H,0,3]$ direction at different temperatures for the $x=0.092$ sample using the two-axis mode and 60 K scattering data
as background.  Figure 2(b) shows similar scans using the triple-axis mode with $\Delta E=0.1$ meV.
The temperature dependence of the magnetic order parameters is shown in Fig. 2(c).  Since the scattering gradually increases with decreasing temperature, it is difficult to determine precisely the
$T_N$ of the system. Nevertheless, we can find its relative changes by using the same
criteria for $T_N$ in both measurements.
From a simple extrapolation of the low and high temperature AF order parameters in Fig. 2(c), we see a clear reduction in the observed $T_N$ on changing
from the two-axis to triple-axis mode. In principle, such a reduction in $T_N$ may result from the temperature 
differences in the measured critical scattering regimes using different instrumental resolutions, 
as the critical scattering temperature regime depends sensitively on the spatial and order parameter dimensionality and is generally large in quasi-two dimensional magnets \cite{birgeneau}.
However, the spin-spin correlation length should still diverge below $T_N$ \cite{birgeneau}.
Figure 2(d) shows temperature dependence of the spin-spin correlation length, obtained by Fourier transform of the
scattering profiles in Fig. 2(a).  Consistent with the earlier work \cite{hqluo}, we find that the spin-spin correlation length does not diverge
and only reaches to $\sim$50 \AA\ in the low temperature AF ordered state.  These results suggest that the change in the 
observed $T_N$ cannot be due to the effect of critical scattering.

\begin{figure}[t]
\includegraphics[scale=.55]{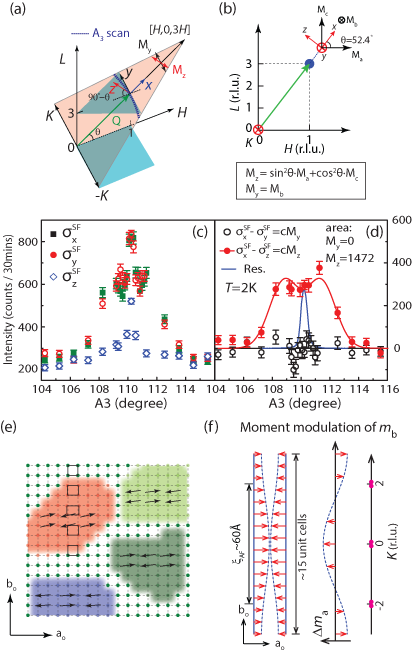}
\caption{(a) Scattering plane (light pink area)
in reciprocal space, $A3$ scan trajectory (blue dashed line),  and the neutron polarization directions ($x, y, z$)
with respective to the wave vector ${\bf Q}$. (b) The relationship between the neutron polarization directions and magnetic moments
along the $a_o$- ($M_a$), $b_o$- ($M_b$), and $c$-axis ($M_c$) directions \cite{hqluo13}.
$M_x$, $M_y$, and $M_z$ are magnetic components along the neutron polarization directions $x$, $y$, and $z$ directions, respectively.
(c) Spin-flip scattering cross sections $\sigma^{\rm SF}_{x}$, $\sigma^{\rm SF}_{y}$, and $\sigma^{\rm SF}_{z}$
in the $A3$ (rocking curve) scans across ${\bf Q}=(1, 0, 3)$ at $T=2$ K. (d) $cM_z \sim M\rm(1, 0, 1/3\rm)$ (solid red line) and $cM_y\sim M_b$ (open circles).  The blue solid line shows the instrument resolution obtained using $\lambda/2$. (e) Schematic of the cluster spin glass in the matrix of the superconducting phase. (f) A model of the moment modulating
spin-density-wave which can give the incommensurate AF order.
 }
\end{figure}

Figure 3 summarizes similar measurements on the $x=0.092$ sample
using TRISP, which can operate as a normal thermal triple-axis spectrometer with $\Delta E\approx 1$ meV
and a NRSE triple-axis spectrometer with $\Delta E\approx 1\ \mu$eV \cite{trisp,inosov}.  Figure 3(a) compares the
instrumental resolution using the normal and NRSE triple-axis modes.
If the AF order is instrumentation resolution limited in the NRSE measurements, we would expect
to find the width of the quasielastic scattering, $\Gamma$, to be $\sim1\ \mu$eV.  From the temperature dependence of $\Gamma$, we see considerable broadening of the quasielastic
scattering above 30 K [Fig. 3(b)].  This is consistent with
temperature dependence of the magnetic order parameters obtained using the NRSE (red diamonds) modes [Fig. 3(c)]. However, identical measurement using normal triple-axis (blue squares) gives a much higher $T_N$ [Fig. 3(c)].  The large variations in the measured $T_N$,
changing from $T_N=37\pm 3$ K at $\Delta E\approx 1$ meV to $T_N=30\pm 2$ K at $\Delta E\approx 1\ \mu$eV, means that the spins of the system freeze
below 30 K on a time scale of $\tau\sim \hbar/\Delta E\approx 6.6\times 10^{-10}$ s, similar to the dynamics of a typical spin glass \cite{binder}.

Another way to establish the origin of the incommensurate AF phase
 is to determine its ordered moment direction.
In an ideal incommensurate spin-density-wave, there should be either ordered moment or moment modulations 
along the incommensurate AF ordering or the $b_o$-axis direction \cite{fawcett}.
 To see if this is indeed
 the case, we used neutron polarization analysis to determine the ordered moment
 direction in the $x=0.096$ sample on IN22 \cite{hqluo13}, which has essentially the same
incommensurate AF order as that of the $x=0.092$ compound.
For this experiment, the sample is aligned in the $[H,0,3H]$ and $[0,K,0]$ scattering plane.  The incident and outgoing neutron beams are polarized
along the $[H,0,3H]$ ($x$), $[0,K,0]$ ($y$), and $[-3H,0,H]$ ($z$) directions [Fig. 4(a)] \cite{hqluo13}.
Using neutron spin-flip (SF) scattering cross sections
$\sigma^{\rm SF}_{\alpha}$, where $\alpha=x,y,z$, we can extract the magnetic moments $M_y$ and $M_z$ via
$cM_y= \sigma^{\rm SF}_{x}-\sigma^{\rm SF}_{y}$ and $cM_z= \sigma^{\rm SF}_{x}-\sigma^{\rm SF}_{z}$ where
$c=(R-1)/(R+1)$ and $R$ ($\approx 15$) is the flipping ratio \cite{hqluo13}. Since the magnetic moment $M_b$ equals to $M_y$ and $M_z=M_a\sin^2\theta +M_c\cos^2\theta$ [Fig. 4(b)], we can conclusively determine $M_b$
by measuring $\sigma^{\rm SF}_{x}$ and $\sigma^{\rm SF}_{y}$. Figure 4(c) shows rocking curve scans
through ${\bf Q_{AF}}=(1,0,3)$ for $\sigma^{\rm SF}_{\alpha}$.
The estimated $M_b\sim cM_y$ and $M_z\sim c(M_a\sin^2\theta +M_c\cos^2\theta)$ are plotted in Fig. 4(d).
To within the errors of our measurements, we find $M_b=0$ meaning no measurable moments along the
$b_o$-axis direction.
Since our data are collected by rotating the crystals at fixed $\left|{\bf Q_{AF}}\right|$, we are effectively measuring the mosaic distribution of the longitudinally ordered AF phase [Fig. 4(e)]. 

\section{Discussion and Conclusions}

To understand the microscopic origin of the incommensurate AF phase, we consider two possibilities as sketched in Figs. 4(e) and 4(f).
If the ordered moments of the incommensurate phase are aligned along the longitudinal ($a_o$-axis) direction, the observed incommensurate scattering
may be the mosaic distribution of the commensurate AF phase in the matrix of the superconducting phase as shown in the green patches 
of Fig. 4(e).  However, a mosaic distribution of the commensurate AF phase should result in a broad peak centered at the AF wave vector, in contrast to the observed transverse incommensurate AF order.  Furthermore, such a model cannot explain the magnitude or the doping dependence of 
the incommensurability.  It also does not account for the expected orthorhombic lattice distortion in the incommensurate AF phase. Therefore, this model is unlikely to be correct description of the observsed incommensurate phase.
Alternatively, if the incommensurate AF order arises from the moment amplitude modulation along the $b_o$-axis [Fig. 4(f)], 
an incommensurability of
$\delta=0.03$ would require a spin-spin correlation length of $\sim$15 unit cells or $\sim$80 \AA, only slightly larger than the 
observed $\sim$60 \AA\ correlation length.  In principle, an $a_o$-axis aligned moment of 
the incommensurate AF order cannot exist in the tetragonal unit cell and must break the $C_4$ rotational 
symmetry of the underlying tetragonal 
crystalline lattice [left inset in Fig. 1(a)] \cite{qhuang}.  When incommensurate AF order is initially established below $T_N$, one can observe clear orthorhombic lattice distortion in X-ray diffraction experiments \cite{xylu}.  When superconductivity sets in below $T_c$, the orthorhombic lattice distortion and intensity of incommensurate AF peaks are gradually suppressed with decreasing temperature \cite{nandi,pratt2,hqluo,xylu}. This is consistent with the picture that 
incommensurate AF order is intimately associated with the orthorhombic lattice distortion.
Assuming optimal superconductivity in BaFe$_{2-x}$Ni$_{x}$As$_{2}$ prefers a true tetragonal 
structure, this means that the incommensurate AF phase must
 be located in the 
 orthorhombic lattice distorted patches in the matrix of the paramagnetic tetragonal phase below $T_N$ [red region in Fig. 4(e)].  
Since the lattice distortion from the tetragonal
 to orthorhombic phase must be gradual, one can imagine a scenario where the incommensurate AF order arises from the 
 moment amplitude modulation 
  along the $b_o$-axis coupled with the tetragonal to orthorhombic structural transition [Red Patch in Fig. 4(e) and Fig. 4(f)].
Below $T_c$, the volume fraction of the superconducting tetragonal phase grows with decreasing temperature 
the expense of the incommenusrate AF orthorhombic phase. 				
Although we cannot 
conclusively determine whether this picture is correct, it is consistent  
with a cluster spin glass (or amplitude spin glass) and much different from a spin-density wave in nested Fermi surfaces.

The identification that the short-range incommensurate AF phase, a general feature in electron-doped iron pnictides \cite{pratt2,hqluo}, 
is a cluster spin glass challenges the notion that the static AF order in iron pnictides arises from the itinerant electrons in
nested Fermi surfaces. 
Furthermore, since incommensurate AF order competes directly with superconductivity \cite{pratt2,hqluo},
one can envision a situation where the cluster spin glass coexists and competes
mesoscopically with superconductivity near optimal electron-doping.
While these results are consistent with $\mu$SR measurements on BaFe$_{2-x}$Ni$_{x}$As$_{2}$ indicating
that the disappearance of static magnetism with increasing $x$ is driven mainly by the loss of the volume
fraction of the magnetically ordered region near optimal superconductivity \cite{arguello},
they are different from the underdoped regime where antiferromagnetism and superconductivity coexist
homogeneously and compete for the same itinerant
electrons \cite{pratt1,Christianson,mywang,fernandes1,fernandes2}.  These results are also consistent with $^{57}$Fe
M$\rm \ddot{o}$ssbauer spectroscopy measurements on the same Ni-doped BaFe$_{2-x}$Ni$_{x}$As$_{2}$ \cite{arguello,munevar}.
With increasing electron-doping, the long-range commensurate AF order transforms into a cluster spin glass via the first order fashion.  Upon further doping,
the cluster spin glass is replaced by a homogeneous superconducting phase with tetragonal structure.
The behavior of the incommensurate AF 
ordered phase in BaFe$_{2-x}$Ni$_{x}$As$_{2}$ is remarkably similar to those of the 
hole underdoped copper oxides superconductors such as La$_{1.94}$Sr$_{0.06}$CuO$_4$ \cite{sternlieb} and YBa$_2$Cu$_3$O$_{6+x}$ \cite{haug}, 
where the spin freezing temperature depends sensitively 
on the energy resolution of the probes.  Very recently, incommensurate charge ordering 
in underdoped copper oxides was also found to compete with superconductivity \cite{Ghiringhelli,Tacon}.  These results, together with the present finding 
of a cluster spin glass phase in iron pnictides, suggest that the spin and charge ordering competing with superconductivity 
may be a general phenomenon in the phase diagram of doped high-transition temperature 
superconductors.

The work at IOP, CAS, is
supported by MOST (973 projects: 2012CB821400 and 2011CBA00110) and NSFC (Projects: 11374011 and 91221303). X. L. and H. L. acknowledge Project No. 2013DB03 supported by NPL, CAEP. 
We also acknowledge support from the U.S. NSF-DMR-1308603 (Rice and Oak Ridge) and 
the Robert A. Welch Foundation Grant No. C-1839 at Rice.


\end{document}